\documentclass[epsf]{elsart}
\usepackage{graphicx,amsmath,epsf}
 \let\oldensuremath=\ensuremath\renewcommand\ensuremath{\relax\oldensuremath}

\newcommand{\be}{\begin{equation}}
\newcommand{\ee}{\end{equation}}
\newcommand{\bea}{\begin{eqnarray}}
\newcommand{\eea}{\end{eqnarray}}

\newcommand{\gsim}{\ensuremath{\mathrel{\raise2pt\hbox to 8pt{%
                        \raise -5pt\hbox{$\sim$}\hss{$>$}}}}}

\def\vev#1{\langle      #1 \rangle}

\def\eqinf{\mathrel{\raise2pt\hbox to 24pt{\raise -8pt\hbox{$t \to \infty$}\hss{$=$}}}}

\def\gsim{\mathrel{\raise2pt\hbox to 8pt{\raise -5pt\hbox{$\sim$}\hss{$>$}}}}
\def\rsim{\mathrel{\raise2pt\hbox to 8pt{\raise -5pt\hbox{$\sim$}\hss{$>$}}}}
\def\lsim{\mathrel{\raise2pt\hbox to 8pt{\raise -5pt\hbox{$\sim$}\hss{$<$}}}}

 \newif\ifchi\newif\iflat\newif\ifMeV
 \newif\iflight\newif\ifinter\newif\ifstrange
 \newif\ifK\newif\ifKstar\newif\ifphi
 \chifalse\lattrue\MeVtrue \lighttrue\intertrue\strangetrue
 \Ktrue\Kstartrue\phifalse

 \newcount\colcount\newcount\allcolcount
 \newcommand\setupcounts{%
    \colcount=0
    \iflight  \advance\colcount by 1\fi
    \ifinter  \ifK    \advance\colcount by 1\fi
              \ifKstar\advance\colcount by 1\fi
              \ifphi  \advance\colcount by 1\fi\fi
    \ifstrange\ifK    \advance\colcount by 1\fi
              \ifKstar\advance\colcount by 1\fi
              \ifphi  \advance\colcount by 1\fi\fi
    \allcolcount=1
    \ifchi\advance\allcolcount by 1\fi
    \iflat\advance\allcolcount by \colcount\fi
    \ifMeV\advance\allcolcount by \colcount\fi
    \ifcase\colcount\begingroup\errhelp{can't switch off all columns}%
                               \errmessage{colcount is 0}\endgroup\fi
 }
\input{epsf}
\begin{document}

   \begin{flushright}
   \vspace{-0.3cm}
   \small{LAUR-99-789} \\
   \end{flushright}

\begin{frontmatter}
  
  \title{General Physics Motivations for Numerical Simulations of
    Quantum Field Theory}
  
  \author{ R. Gupta\thanksref{email}}
  
  \address{Theoretical Division, Los Alamos National Laboratory, Los
    Alamos, NM 87545, USA}
  
  \thanks[email]{\tt rajan@lanl.gov}
  
  \date{VERSION --- \today}

\begin{abstract}
\noindent
In this introductory article a brief description of Quantum Field
Theories (QFT) is presented with emphasis on the distinction between
strongly and weakly coupled theories.  A case is made for using
numerical simulations to solve QCD, the regnant theory describing the
interactions between quarks and gluons.  I present an overview of what
these calculations involve, why they are hard, and why they are tailor
made for parallel computers. Finally, I try to communicate the
excitement amongst the practitioners by giving examples of the
quantities we will be able to calculate to within a few percent
accuracy in the next five years.

%PACS numbers: 12.38.Gc, 12.39.Hg, 13.20.He, 14.40.Nd
\end{abstract}


\end{frontmatter}

\section{Introduction}
\label{sec:intro}
The complexity of physical problems increases with the number of
degrees of freedom, and with the details of the interactions between
them.  This increase in the number of degrees of freedom introduces
the notion of a very large range of length scales.  A simple
illustration is water in the oceans.  The basic degrees of freedom are
the water molecules ($\sim 10^{-8}$ cm), and the largest scale is the
earth's diameter ($\sim 10^{4}$ km), $i.e.$ the range of scales span
$10^{17}$ orders of magnitude. Waves in the ocean cover the range from
centimeters, to meters, to currents that persist for thousands of
kilometers, and finally to tides that are global in extent. Solutions
to modeling the behavior of ocean waves depends on whether this whole
range of scales are important or whether one can isolate them. For
example, to understand the currents and tides one does not need to
know the molecular nature of water at all, one can treat it as a
continuous fluid. In general, different techniques are necessary
depending on whether disparate length scales are strongly or weakly
coupled.

The study of interactions between elementary particles involves length
scales ($ < 10^{-13}$ cm ) where quantum effects are large.  The
quantum vacuum is no longer a smooth placid entity, but more like a
frothing bubbling pot of stew.  It has fluctuations at all scales from
$ 10^{-13}$ cm to zero.  The mathematical framework developed over the
last fifty years for handling these fluctuations at all length scale
is called Quantum Field Theory (QFT). (For a background on QFT see 
the text books~\cite{QCDbooks}.) Once again, the techniques used
for solving these theories depend on how strongly these vacuum
fluctuations couple to elementary processes like the scattering of two
particles. These ideas are developed in Section~\ref{sec:QFT}. The
coupling between quarks and gluons, the strongly interacting
elementary particles, is $O(1)$ at the hadronic scale $ 10^{-13}$ cm.
The associated QFT, called Quantum Chromodynamics (QCD), therefore
involves both an infinite range of energy scales and a strong
coupling. Thus, to answer many interesting questions like how neutrons
and protons arise as bound states of quarks and gluons necessitates
techniques different from the usual $--$ series expansion in a small
coupling, $i.e.$ perturbation theory.  The set of articles in this
book explore the technique based on the numerical simulations of the
underlying QFT discretized on a finite space-time grid.  This brute
force approach is the only known technique that is in principle exact.
The remainder of this article is an attempt to show how the power of
today's supercomputers, especially massively parallel computers, can
be used to solve this highly non-linear, strongly interacting QFT.

\section{What is Quantum Field Theory}
\label{sec:QFT}
A pervasive feature of collisions between very high energy particles
is the lack of conservation of the number of particles. The
annihilation of an electron and positron, say each of energy 1 TeV in
the center of mass, produces, in general, hundreds of particles
(electrons, positrons, muons, pions, $etc.$). Thus, a theory of such
interactions has to allow for the possibility of the spontaneous
creation and annihilation of the various particles observed in nature.
Such a theory should also obey the rules of quantum mechanics and
relativity (Lorentz invariance). Quantum field theories have
been remarkably successful in explaining three of the four forces:
strong (QCD), electromagnetic(QED), and weak(WI).  The effort of high
energy physicists today is to develop techniques to extract very
precise predictions of these theories and to confront these against
results of experiments to determine if the theories are complete.  The
set of articles in this book explore a technique based on numerical
simulations of QFT. Other physicists are trying to incorporate
gravitational interactions, for which QFT has not been very
successful, into our overall understanding.  This effort has lead to
exciting developments in string/M theory, however the discussion of
these is beyond the scope of this article.

In QFT, the basic degrees of freedom are fields of operators defined
at each and every space time point. These variables satisfy
appropriate commutation/anti-commutation relations depending on their
spin.  Particles emerge as excitations of these field variables, and
can be annihilated and created at any point. In fact there is a
fundamental duality between the description of nature in terms of
particles or fields.  The interactions between the fields is local,
$i.e.$ the equations of motion depend only on the value of the field
variable and derivatives at the same point.  A consequence of this
locality is the causal propagation of particles, $i.e.$ there is no
action at a distance.  Forces between particles are described via the
exchange of intermediary fields/particles like the photons, gluons,
$W^{\pm}$, and $Z$ bosons.

When a very high energy electron and positron annihilate, energy,
momentum, and angular momentum (these are three quantities that are
conserved in all processes) are transferred to the photon field.  The
resulting excitations in the photon field propagate according to its
equations of motion, and also cause excitations in other variables via
highly non-linear interactions.  Since there is a field variable for
each kind of elementary particle, one can account for processes in
which one kind of particle is annihilated and another is created.
This produces a cascade of particles, which if captured in the very
sophisticated detectors used in modern experiments, give a readout of
their properties -- charge, energy, and momentum.  The process
is intrinsically quantum mechanical, $i.e.$ each possible outcome
consistent with the constraints of conservation of energy, momentum,
angular momentum, and other internal symmetries like baryon and lepton
number, has a non-zero probability of occurrence.  For example, in the
$e^+e^-$ annihilation, the end result could be billions of very low
energy photons.  Including such a possibility leads us to the amazing
notion of infinitely many degrees of freedom. Miraculously, this
feature is built into QFT, the field can have many excitations at
every space-time point, and thus infinitely many if there is enough
energy to excite them.

With this very brief introduction to QFT, I can now spring the million
dollar question: if quantum field theories are the correct
mathematical constructs for describing nature, and if these, as
emphasized above, have infinitely many variables all of which play a
role in predicting the outcome of a process, then how can one hope to
study them on a computer which, in finite time, can only do a finite
number of computations involving a finite number of variables?
Clearly, we must have a reasonable way to make the problem finite, for
otherwise this book would end here. The next sections discuss this
process of discretization as well as motivating the need for numerical
methods.

\section{Motivating numerical methods for QFT}
\label{sec:numQFT}

\begin{figure}[th] %1
  \vspace{9pt} \hbox{\hskip15bp\epsfxsize=0.9\hsize \epsfbox
    {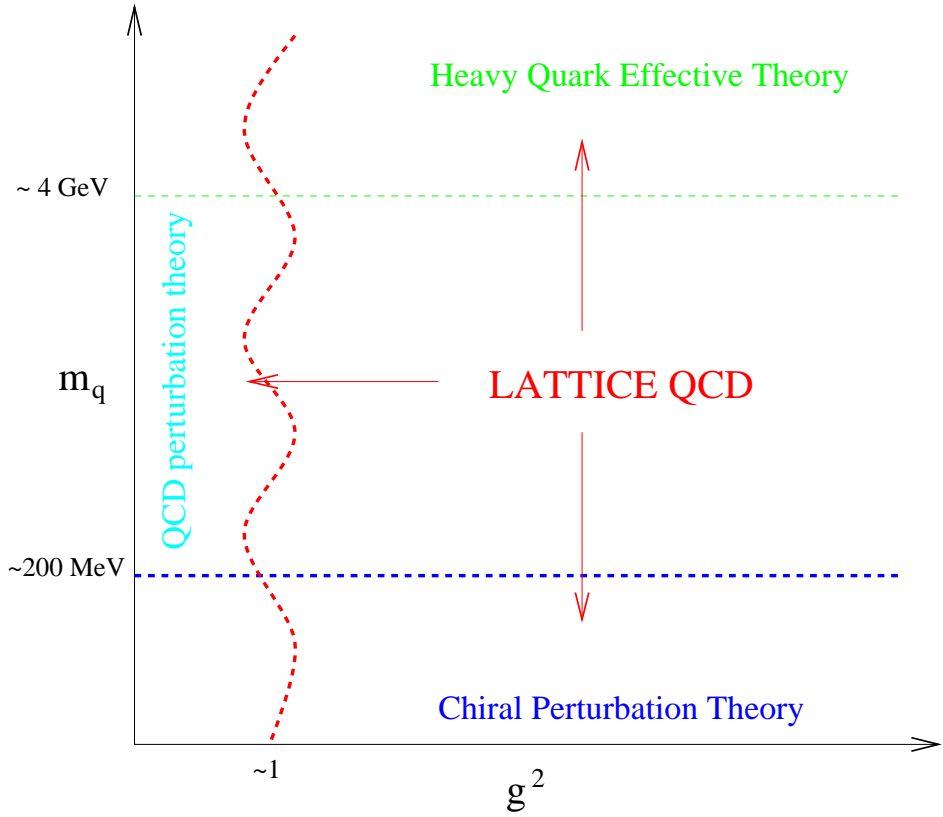}}
\caption{A schematic of the theoretical approaches used to analyze QCD
  and their domain of reliability in terms of the basic parameters,
  the strong coupling constant $g$ and quark masses $m_q$.}
\label{f:QCDdomains}
\end{figure}

In the absence of closed form analytical solutions of quantum field
theories describing nature, one has to investigate different
approximation schemes.  The approximation schemes can be of two kinds:
(a) First principles: retain the basic degrees of freedom but
approximate the calculations. This approach includes the standard
perturbative expansion in the coupling $\alpha$ utilizing Feynman
diagrams, and numerical simulations of QFT discretized on a lattice.
(b) Models: formulate effective degrees of freedom and interactions
between them that retain as many features of the original theory as
possible.  An example in QCD is chiral perturbation theory which has
been successful in describing the low energy nuclear interactions of
pions~\cite{Pich}. In this approach one replaces the quarks and gluons
of QCD by pions as the basic degrees of freedom.  The interactions
between pions have the same symmetries as QCD but involve
phenomenological parameters.  Each of the above approaches have their
domain of reliability and success as illustrated in
Fig.~\ref{f:QCDdomains}.  Our understanding of QCD has relied on input
from all these approaches.  LQCD calculations are unique in the sense
that they can provide the bridge between different approaches and
validate the success of the models in their respective domains.  The
reason is that we can dial the input parameters (coupling and quark
masses) and study the behavior of the theory as a function of these in
the different domains.

\begin{figure}[th] %1
  \vspace{9pt} \hbox{\hskip15bp\epsfxsize=0.9\hsize \epsfbox
    {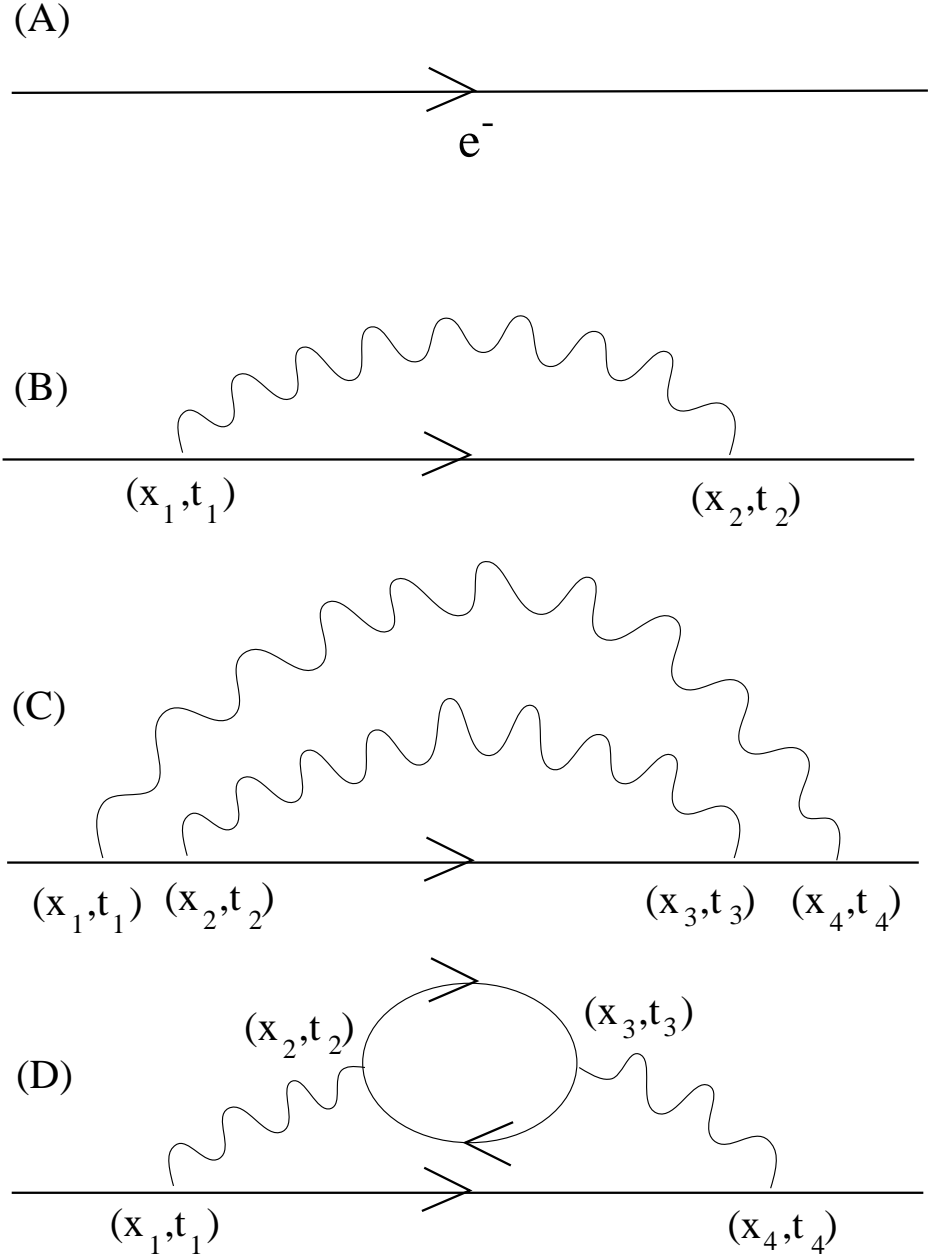}}
\caption{A schematic of the virtual processes, (B) to (D),
  associated with the propagation of a `bare' electron, (A) }
\label{f:electron}
\end{figure}

\begin{figure}[th] %1
  \vspace{9pt} \hbox{\hskip15bp\epsfxsize=0.9\hsize \epsfbox
    {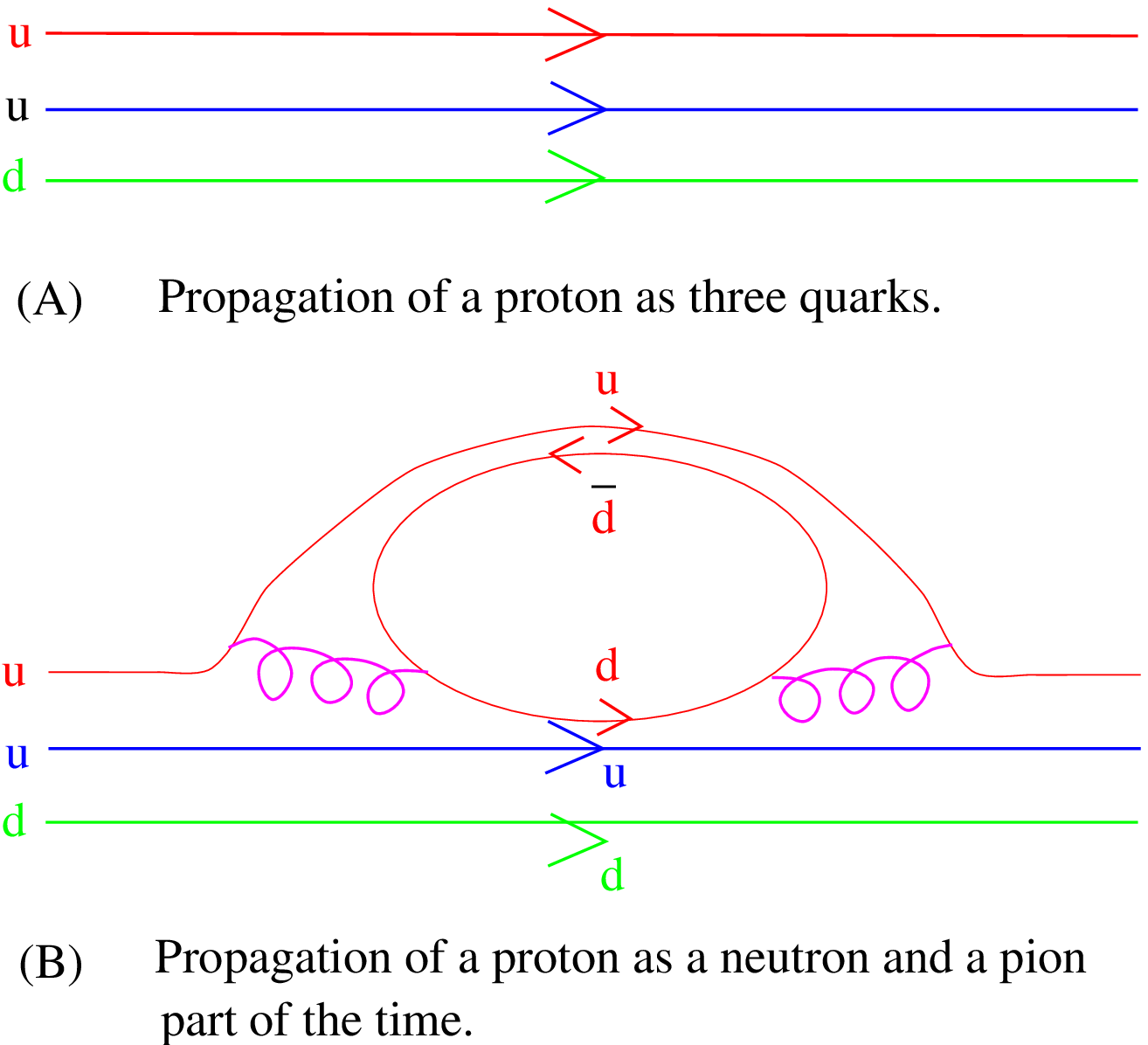}}
\caption{A schematic of the propagation of a proton. Part of the time it
  is in the intermediate virtual state of a neutron and a pion.}
\label{f:proton}
\end{figure}

To appreciate the difference between weakly and strongly interacting
particles, consider the propagation of an electron and a proton as
illustrated in Figs.~\ref{f:electron} and \ref{f:proton}. The quantum
mechanical wavefunction of the electron (called the Fock space
wavefunction) is a superposition of the electron propagating
undisturbed; the electron propagating by emitting a photon at point
$x_1,t_1$ and reabsorbing it at point $x_2, t_2$; emitting two photons
at points $x_1,t_1$ and $x_2,t_2$ and reabsorbing them at points
$x_4,t_4$ and $x_3, t_3$; emitting a photon at point $x_1,t_1$, the
photon generating an electron-positron pair at $x_2, t_2$, which
annihilate to produce a photon at $x_3, t_3$, which is reabsorbed by
the initial electron at $x_4,t_4$; and so on.  These higher order
processes are suppressed by the number of times the interaction
between the photon and charge takes place. For example, in
Fig.~\ref{f:electron}D, the suppression factor for the amplitude is
$g_{em}^4$ ($\alpha = g_{em}^2/4\pi \approx 1/137$).
%% probability ($|amplitude|^2$) for that subprocess $e^+ + \gamma \to e^+ e^+ e^-$ is very
%%small, $\propto \alpha_{em}^2 = (1/137)^3 \approx 6 \times 10^{-4}$. 

Furthermore, these higher order process are ``virtual''; $i.e.$ an
electron in an eigenstate does not change unless one tries to detect
it or there is an interaction with an external source.  The very
process of detection disturbs the electron and imparts it energy and
momentum.  Thus, the electron manifests itself as two electrons and a
positron a very small fraction of the time unless the probe imparts a
lot of energy.

The bottom line is that if the coupling constant associated with a
given kind of interaction is small, as is true of electromagnetic and
weak interactions, the simplest possibility is dominant and the higher
order possibilities are small corrections.  These corrections can be
calculated analytically as a power series expansion in the coupling.
The most striking and successful example of this approach is the
calculation of the anomalous magnetic moment of the electron $a_e =
0.5(g_e-2)$. Kinoshita and collaborators~\cite{Kinoshita96} calculated
891 4-loop diagrams (Fig.~\ref{f:electron}C,D are examples of 2-loop
diagrams), and find
\begin{eqnarray}
a_e(theory) &=& 0.5 \big( \alpha / \pi \big) + \nonumber \\
            &-& 0.328\ 478\ 965\ \ldots \big( \alpha / \pi \big)^2  \nonumber \\
            &+& 1.181\ 241\ 456\ \ldots \big( \alpha / \pi \big)^3  \nonumber \\
            &-& 1.409\ 2\ (384)\ \ldots \big( \alpha / \pi \big)^4  \nonumber \\
            &+& 4.396\ (42)\ \times 10^{-12}, 
\label{eq:kinoshitath}
\end{eqnarray}
to be compared with the experimental measurements
\begin{eqnarray}
a_{e^-}(exp) &=& 1\ 159\ 652\ 188.4\ (4.3)\ \times 10^{-12} \nonumber \\
a_{e^+}(exp) &=& 1\ 159\ 652\ 187.9\ (4.3)\ \times 10^{-12} \,.
\label{eq:kinoshitaEXP}
\end{eqnarray}
To confront theoretical prediction with experimental numbers one needs
an independent measurement of the electromagnetic coupling $\alpha$.
It turns out that the uncertainty in the measurements of $\alpha$ from
the quantum Hall effect, the ac Josephson effect, and the de Broglie
wavelength of the neutron beam, are large enough to deny a resolution
of this issue!  So in fact the authors have turned the question around
$--$ assume QED is correct and use $a_e$ to determine $\alpha$!  This
example should convince you that very precise predictions of QED can
be extracted using perturbation theory; the exception being when
many-body effects are important as for example in some condensed
matter systems. The same is true of weak interactions as these are
governed by an even smaller coupling. Parenthetically it is worth
highlighting that this ``analytical'' calculation~\cite{Kinoshita96}
could not have been done without the resources of a massively parallel
computer.

Before discussing QCD, it is worth mentioning that the coupling
constant in QFT is not a fixed constant. Quantum fluctuations in the
form of virtual particles discussed above can screen or anti-screen
the charge.  As a result, the strength of interactions depends on the
energy of the probe. In QCD the coupling is large, $O(1)$, for energy
scales $\lsim 1 $ GeV (typical of nuclear matter), and becomes weak
with increasing energy. This is why perturbative methods are reliable
for describing high energy collisions (small $g$ region in
Fig.~\ref{f:QCDdomains}), while non-perturbative methods like lattice
QCD are essential at nuclear energy scales.

In Fig.~\ref{f:proton}B I have shown the virtual process in which the
intermediate state of a proton is a neutron and a pion. Once again the
leading contribution is $g_s^4$, however since $\alpha_s = g_s^2/4\pi
\sim 1$, this and other such ``higher'' order processes are not
suppressed.  Thus the situation in the case of strong interactions,
described by QCD, is completely different. QCD is a highly non-linear,
strongly interacting theory characterized by $\alpha_s \sim 1$ for
processes involving momentum transfers of $\lsim 1$ GeV. As a result
perturbative calculations based on keeping a few low order Feynman
diagrams are not reliable.  One has to include essentially all
exchanges of soft gluons, $i.e.$ devise non-perturbative approaches.
Large scale numerical simulations of lattice QCD are currently the
most promising possibility; in this approach contributions of all
possible interactions via the exchange of gluons with momentum smaller
than the lattice cutoff $2 \pi /a$ are included.

\section{QCD}
\label{sec:QCD}
Deep inelastic scattering experiments, scattering of very high energy
electrons off protons and neutrons, carried out in the late sixties
and early seventies at SLAC showed that the latter were composite.
They were bound states of even more fundamental particles which were
given the name quarks by Gell-Mann (or aces by Zweig). These quarks
carried, in addition to the electromagnetic charge $e/3$ or $2e/3$, a
strong charge that was dubbed color.  If QFT were to explain the
experiments, three generalizations to QED were required.  First, the
color charge had to occur in three varieties (which were called red,
blue, and green) but each with the same strength.  Second, the value of
this charge had to depend on the energy, it was strong at low energies
and progressively got weaker as the energy was increased.  Third, the
quarks were always confined within hadrons as no experiment had
detected an isolated quark. It turned out that a generalization of QED
with precisely these additional features had been proposed, in a
different context, in 1954 by Yang and Mills~\cite{yangMills}.  This
mathematical theory adapted to the features seen in strong
interactions is called QCD.

In QCD, quarks can carry any one of the three colors and anti-quarks
carry anti-color.  The interactions between them is via the exchange
of gluons which themselves carry color.  Thus, a red quark can turn
into a green quark by the emission of a red-antigreen gluon.
Similarly a green quark and an anti-green anti-quark can annihilate
into a green anti-green gluon or a photon which has no strong charge.
The fact that gluons, unlike photons, also carry the color charge
makes QCD a much more non-linear theory than QED.

One way confinement can arise is if the potential energy between a
quark and antiquark (or two quarks) grows linearly with distance,
consequently it would take an infinite amount of energy to isolate
them. The hadrons we detect in experiments would have to be color
neutral. There are two types of color neutral objects one can form
with this three color theory, mesons made up of a quark and an
antiquark with the corresponding anti-color, and baryons with three
quarks having different colors but in a color neutral combination.
The mathematical structure that encapsulates these ideas is the
non-abelian group of $3\times 3$ unitary matrices with unit
determinant, SU(3). The quarks are assigned to the fundamental
representation {\bf 3} of SU(3) and anti-quarks to the $\bf {\bar 3}$.
The color neutral mesons and baryons are the combinations that
transform as singlets, $i.e.$ the identity representation in $3
\otimes \bar{3} = 1 \oplus 8$ and $3 \otimes 3 \otimes 3 = 1 \oplus 8
\oplus 8 \oplus 10$ respectively.  For a more detailed introduction to
QFT and QCD see~\cite{QCDbooks}.

To explain the plethora of hadrons that had already been seen in
experiments by the early sixties, it was postulated that quarks have
yet another label (internal quantum number) called flavor. As of
today, experiments have revealed six flavors, which are called up,
down, strange, charm, beauty, and top, and the strong interactions are
observed to be the same for all. This replication in nature of quarks
with flavor is very similar to that seen in the case of leptons;
electrons, muons, and tau have identical electromagnetic interactions
and differ only in their mass.  There is, however, one major
difference.  We need at least two flavors, up and down quarks, to form
neutrons and protons.  These, along with electrons, are the building
blocks of all matter.

All these features of strong interactions are neatly summarized in the
Yang-Mills action density
\begin{equation}
S(x) = \sum_{flavors}^{i} \ \sum_{colors}^{a,b}  \overline{\psi}^{i,a}(x) 
\big[ i \gamma_\mu ( \partial_\mu \delta^{ab}- i g G_\mu^{a,b}) -m_i \big] \psi^{i,b}
\label{eq:dirac}
\end{equation}
where $G_\mu = 1/\sqrt{2} \sum_i T^i A_\mu^i$ is the matrix of gauge
fields and the $T^i$ are the eight generators of SU(3). The action for
the gauge fields is
\begin{eqnarray}
S_{gauge}(x) &=& -1/2 {\rm Tr} F^{\mu\nu}F_{\mu\nu} \nonumber \\
F_{\mu\nu}^a &=& \partial_\mu G_\nu^a - \partial_\nu G_\mu^a + g f^{abc}G^b_\mu G^b_\nu \,.
\label{eq:gauge}
\end{eqnarray}
This action has an exact invariance under local (independent at each
and every space-time point) transformations in the color variables,
again a generalization of the gauge transformation of QED.  The only
differences from the Dirac action for QED are (i) the color indices in
Eqs.~\ref{eq:dirac} and \ref{eq:gauge}, and (ii) the non-linear term
in Eq.~\ref{eq:gauge} which arises because the gluons themselves carry
charge and are self-interacting. Thus, for small $g$ one can imagine
that this non-linear term will be small and a perturbative expansion
in $g$ is reliable.  This is exactly what happens at very high
momentum transfers.  On the other hand, when $g$ is large the
non-linear effects cannot be neglected and one needs different
non-perturbative methods. The method we will discuss is 
numerical simulations of QCD discretized on a space-time lattice.

\section{Overview of the Lattice Approach}
\label{s:overviewLQCD}

LQCD calculations are a non-perturbative implementation of field
theory using the Feynman path integral approach. The calculations
proceed exactly as if the field theory was being solved analytically
had we the ability to do the calculations.  The starting point is the
partition function in Euclidean space-time
\begin{equation}
Z \ = \
\int \big[ \prod_x {\mathcal D}A_{\mu}(x) \ {\mathcal D}\psi(x) \ {\mathcal D}
\bar \psi(x) \big] \  e^{-S}
\end{equation}
where $S$ is the QCD action
\begin{equation}
{\mathcal S} \ = \ \int d^4 x \ \big( \frac{1}{4} F_{\mu\nu}F^{\mu\nu}(x) -
                            \bar \psi M \psi \big) \ .
\end{equation}
Here $F_{\mu\nu}F^{\mu\nu}(x)$ is the gluon action density and and $M$
is the Dirac operator. The fermions are represented by Grassmann
variables $\psi$ and $\bar \psi$. Since the action is linear in $\psi$
and $\bar \psi$, these can be integrated out exactly using rules of
Grassman integration with the result
\begin{eqnarray}
Z \ &=& \ \int \big[ \prod_x {\mathcal D}A_{\mu}(x) \big] \ {\rm det}M\ 
          e^{\int d^4 y \ (-\frac{1}{4} F_{\mu\nu}F^{\mu\nu})} \nonumber\\
    &=& \ \int \big[ \prod_x {\mathcal D}A_{\mu}(x) \big] \ 
          e^{\int d^4 y \ (-\frac{1}{4} F_{\mu\nu}F^{\mu\nu} 
                                  + \sum_i {\log} ( {\rm Det} M_i))} .
\label{eq:pathintegral}
\end{eqnarray}
The second term defines an effective action for the gauge fields 
\begin{equation}
S_{eff} \ = {S_{gauge} + S_{quarks}} = \ \int d^4 x \ \big( {1/4}
F_{\mu\nu}F^{\mu\nu} \big) - \sum_i {\log} ( {\rm Det} M_i) \,.
\end{equation}
The fermionic contribution is now contained in the highly non-local
term ${\rm det}M $ which is only a function of the gauge fields. 
The sum is over the quark flavors, distinguished by the value of the
bare quark mass. A key feature of QCD is that the action $S$ is real
and bounded from below.  Thus Eq.~\ref{eq:pathintegral} defines a
partition function in exact analogy to statistical mechanics systems
where $Z= \sum_{configurations} e^{-E/kT}$. The only difference is
that instead of the Boltzmann factor $e^{-E/kT}$ we have $e^{-S}$,
$i.e.$ the action measured in units of the Planck constant $\hbar$
which is set to unity by an appropriate choice of units. In numerical
simulations, this factor is used to define the probability with which
to generate the background gauge configurations.

Having generated an ensemble of gauge configurations, results for
physical observables are obtained by calculating expectation values
\begin{equation}
\vev{{\mathcal O}} \ = \ \frac{1}{Z} \ 
\int \big[ \prod_x {\mathcal D} A_{\mu}(x) \big] \ {\mathcal O} \ e^{-S} \ .
\label{eq:expvalue}
\end{equation}
where ${\mathcal O}$ is any given combination of operators expressed
in terms of time-ordered products of gauge and quark fields.  This
calculation involves a double infinity due to the continuous valued
nature of space-time and the gauge fields $--$ the number of points
$x$ are infinite, and at each point $x$ the field $A_\mu$ takes on a
continuous infinity of values. To make the system finite
Wilson~\cite{Wilson74} made two approximations: (i) he replaced
continuous space time by a discrete lattice of finite extent $L$ and
spacing $a$, and (ii) he devised a method that generated a
representative set of important configurations that allow for a very
good approximation to the integral over $A_\mu$. The latter is
achieved using Monte Carlo methods of integration. On this lattice,
the gauge fields $A_\mu(x)$ are defined in terms of $3 \times 3$
complex matrices $U_\mu(x) = \exp(iagA_\mu(x))$ and associated with
the links between sites $x$ and $x + \hat \mu$. The quarks fields in
$\mathcal {O}$ are, in practice, re-expressed in terms of quark
propagators using Wick's theorem for contracting fields. In this way
all dependence on quarks as dynamical fields is removed. For more
details on the discretization of QCD on the lattice
see~\cite{Creutz83,RajanLH}.

The basic building block for fermionic quantities is the Feynman
propagator,
\begin{equation}
S_F(y,j,b;x,i,a) \ = \ \big( M^{-1} \big)^{y,j,b}_{x,i,a} \, ,
\end{equation}
where $M^{-1}$ is the inverse of the Dirac operator calculated on a
given background field. A given element of this matrix $\big( M^{-1}
\big)^{y,j,b}_{x,i,a}$ is the amplitude for the propagation of a quark
from site $x$ with spin-color $i,a$ to site-spin-color $y,j,b$.  The
matrix $M$ is very large, $12L^4 \times 12L^4 $, but very sparse.  A
given site is connected to only itself and a few neighbors, $i.e.$ a
total of 9 points in the case of nearest-neighbor interactions.  Thus,
in numerical calculations $M$ is not stored but constructed on the fly
from the gauge links $U_\mu$, which define it completely. The
calculation of the inverse is done by solving the set of linear
equations
\begin{equation}
M \chi = \phi
\label{eq:Minverse}
\end{equation}
where $\phi$ is a source vector (usually taken to be a delta function
at one site).  The solution $\chi$ is found using a variety of
iterative algorithms (Krylov solvers) like conjugate gradient, minimal
residue, $etc.$, that include preconditioning and acceleration
techniques as discussed by Th. Lippert~\cite{Lippert}.  In practice we
do not construct the entire inverse but only solve for a few columns
of $M^{-1} $, $i.e.$ solve Eq.~\ref{eq:Minverse} for a few $\phi$, as
each column gives the Feynman propagator from a given site to all
other sites. Since the theory is invariant under translations, the
particular choice of the site does not matter after the sum over 
configurations. 

To summarize, in the Feynman path approach to QFT one needs the
ability to (a) generate all possible configurations of gauge fields
and (b) calculate a variety of correlation functions on each of these
configurations. Expectation values are averages of these correlation
functions weighted by the ``Boltzmann'' factor $e^{-S}$ of each
configuration.  In the next section I will show how physical
observables are extracted from these expectation values.

\begin{figure}[th] %1
  \vspace{9pt} \hbox{\hskip15bp\epsfxsize=0.9\hsize \epsfbox
    {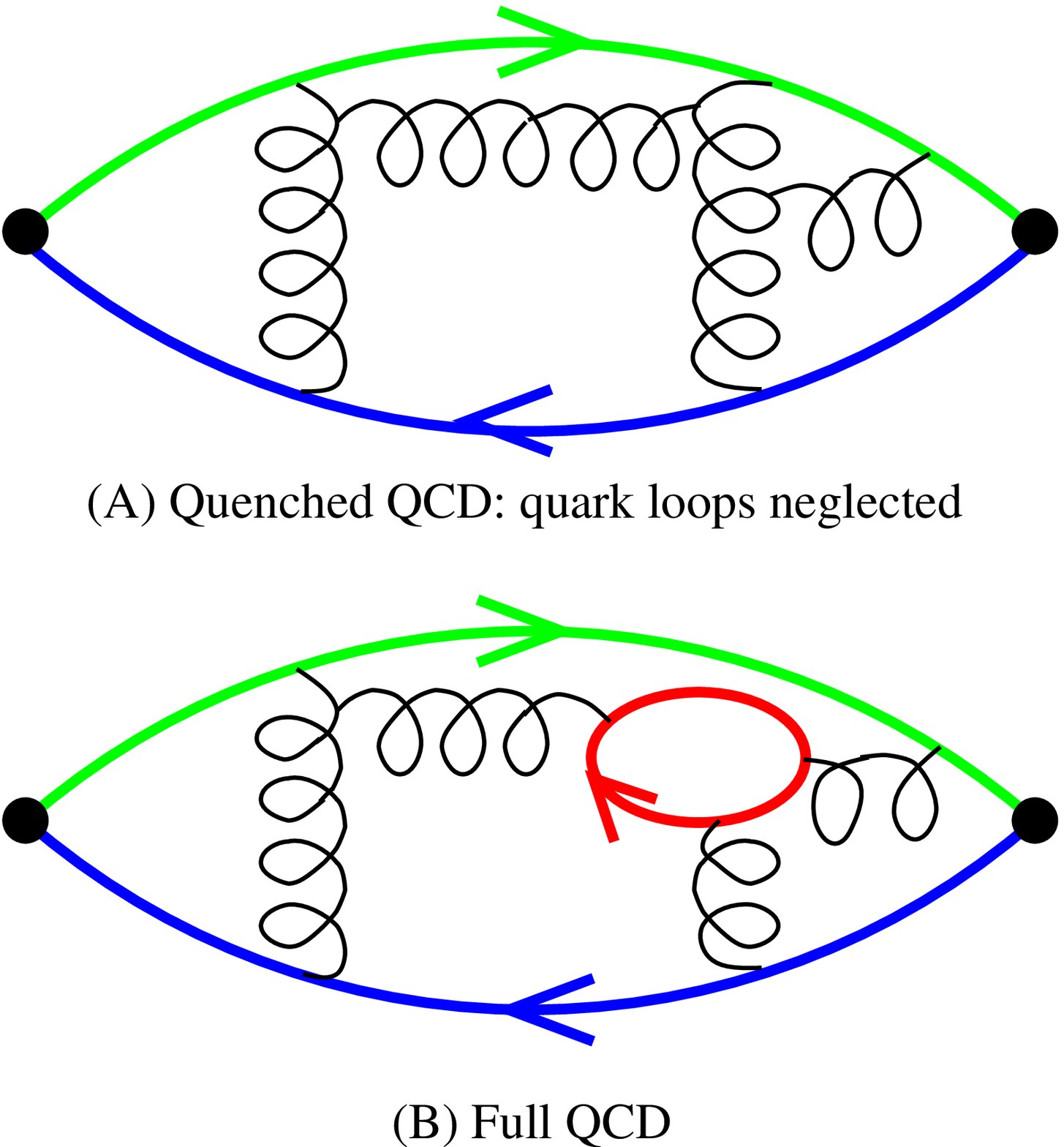}}
\caption{A set of diagrams illustrating the quenched approximation. In it 
  all possible gluon exchanges are allowed but without the possibility
  of internal quark loops (vacuum polarization).}
\label{f:quenched}
\end{figure}

There is a technical issue, the ``quenched'' approximation (QQCD),
worth highlighting here as it has played a central role in
simulations. It consists of setting ${\rm Det} M = constant$ in
Eq.~\ref{eq:pathintegral}~\cite{quenched}.  This approximation
corresponds to removing the momentum dependence of vacuum polarization
effects from the QCD vacuum as illustrated in Fig.~\ref{f:quenched}.
The lattice QCD community has resorted to using this approximation out
of necessity, full QCD computations are $10^3 - 10^5$ times more
costly. The quenched results are, nevertheless, useful
phenomenologically as this approximation is expected to be good to
within $10-20\%$ for many quantities. Also, by doing quenched
calculations we understood all other sources of errors, discussed in
Section~\ref{sec:whyhard}, that are similar in both the quenched and
full theories and how to control them.

\section{Physics from Numerical Simulations}
\label{sec:basic}
In the last section I outlined the calculation of expectation values.
Here I will illustrate how to extract physical observables from these
using as an example the mass $M_\pi$ and the decay constant $f_\pi$ of
the pion.

Consider the 2-point correlation function, $\vev{0 | T[\sum_x {\mathcal
    O}_f(\vec x, t) {\mathcal O}_i(\vec 0,0)] |0 }$ with $t > 0$, where
the operators ${\mathcal O}$ are chosen to be the fourth component of the
axial vector current $ {\mathcal O}_f = {\mathcal O}_i = A_4 = \bar \psi
\gamma_4 \gamma_5 \psi$ as these have a large coupling to the pion.
This 2-point correlation function gives the amplitude for creating a
state with the quantum numbers of the pion out of the vacuum at
space-time point $0$ by the ``source'' operator $ {\mathcal O}_i $; the
evolution of this state to the point $(\vec x,t)$ via the QCD
Hamiltonian; and finally the annihilation by the ``sink'' operator
${\mathcal O}_f $ at $(\vec x,t)$.  The rules of quantum mechanics tell us
that ${\mathcal O}_i$ will create a state that is a linear combination of
all possible eigenstates of the Hamiltonian that have the same quantum
numbers as the pion, $i.e.$ the pion, radial excitations of the pion,
three pions in $J=0$ state, $\ldots$. The second rule is that 
after propagating for Euclidean time $t$, a given
eigenstate with energy $E$ picks up a weight $e^{-Et}$.  Thus, the
2-point function can be written in terms of a sum over all possible
intermediate states
\begin{equation}
\vev{0 | \sum_x {\mathcal O}_f(\vec x,t) {\mathcal O}_i(0) |0 } = \sum_n
\frac{ \vev{0 | {\mathcal O}_f | n} \vev{n| {\mathcal O}_i | 0} }{2 E_n} 
\ e^{-E_n t} \, .
\label{eq:2point}
\end{equation}
To study the properties of the pion at rest we need to isolate this
state from the sum over $n$. To do this, the first simplification is
to use the Fourier projection $\sum_{\vec x}$ as it restricts the sum
over states to just zero-momentum states, so $E_n \to M_n$. (Note
that, by momentum conservation it is sufficient to make the Fourier
projection over either ${\mathcal O}_i$ or ${\mathcal O}_f$.)  The
second step to isolate the pion, $i.e.$ project in the energy of the
state, consists of a combination of two strategies.  One, make a
clever choice of the operators ${\mathcal O}$ to limit the sum over
states to a single state (the ideal choice is to set ${\mathcal O}$
equal to the quantum mechanical wave-functional of the pion), and two,
examine the large $t$ behavior of the 2-point function where only the
contribution of the lowest energy state that couples to ${\mathcal
O}_i$ is significant due to the exponential damping. Then
\begin{equation}
\vev{0 | \sum_x {\mathcal O}_f(x,t) {\mathcal O}_i(0) |0 }
\ \ \eqinf\ \ \ \
\frac{ \vev{0 | {\mathcal O}_f | \pi} \vev{\pi| {\mathcal O}_i | 0}}{2 M_\pi} \ e^{-M_\pi t} \, .
\label{eq:2point1}
\end{equation}
The right hand side is now a function of the two quantities we want
since $ \vev{0 | A_4(\vec p = 0) | \pi} = M_\pi f_\pi$. In this way,
the mass and the decay constant are extracted from the rate of
exponential fall-off in time and from the amplitude.

Let me now illustrate how the left hand side of Eq.~\ref{eq:2point1}
is expressed in terms of the two basic quantities we control in the
path integral -- the gauge fields and the quark propagator.  Using
Wick contractions~\cite{QCDbooks}, the correlation function can be
written in terms of a product of two quark propagators $S_F$, 
\begin{eqnarray}
-\vev{0 | \sum_x \bar \psi (x,t) \gamma_4 \gamma_5 \psi(x,t) &{}& 
\bar \psi (0,0) \gamma_4 \gamma_5 \psi(0,0) |0 } \nonumber \\ 
&{}& \hskip
-0.6 in \equiv \vev{0 | \sum_x S_F(0; \vec x, t) \gamma_4\gamma_5
S_F(\vec x, t; 0) \gamma_4 \gamma_5 | 0 } \,.
\label{eq:2point2}
\end{eqnarray}
and the $\gamma_i$ which are the Dirac spin matrices. 
This is the expectation value we measure as a function of $t$. By
equating its large $t$ behavior to the expected form,
Eq.(\ref{eq:2point1}), we can extract the mass $M_\pi$. The decay
constant $f_\pi$ is obtained from the amplitude using the relation
$\vev{0 | A_4 | \pi} = M_\pi f_\pi$.

Once we have the ability to isolate a given state from the large $t$
behavior there are a number of matrix elements we can calculate from
2-point, 3-point, $\ldots$, correlation functions. Examples are

\begin{itemize}
\item{} $\langle 0 | A_4(\vec p = 0) | \mathcal{P} \rangle$. This gives
  the decay constant for a pseudoscalar hadron $\mathcal{P}$ which can be
  a $\pi$, K, D, or B meson.
\item{} $\langle {\mathcal H}(\vec p, \vec s) | {\mathcal O}_{T}^n | {\mathcal
    H}(\vec p, \vec s) \rangle$.  These matrix elements give the
  $n^{th}$ moment of the momentum distribution of quarks and gluons
  inside the hadron ${\mathcal H}$. Here $T$ labels the ``twist'' of the
  operator. These distributions are measured in deep-inelastic
  scattering experiments.
\item{} $\langle {\mathcal H}(\vec p) | {\mathcal J}_\mu(\vec p) | {\mathcal
    H}(\vec p=0) \rangle$, where ${\mathcal J}_\mu$ is the electromagnetic
  current. These matrix elements probe the charge and current
  distributions within a hadron ${\mathcal H}$ by measuring the
  electromagnetic form-factors.
\item{} $\langle \pi(\vec p) | {\mathcal V}_\mu(\vec p) | {\mathcal B}(\vec
  p=0) \rangle$ and $\langle \pi(\vec p) | {\mathcal A}_\mu(\vec p) |
  {\mathcal B}(\vec p=0) \rangle$ probe the form-factors that parameterize
  the semileptonic decays like $B \to \pi \mu \nu_\mu$.  These are
  used to extract the Cabibbo-Kobayashi-Maskawa matrix elements that
  describe the mixing between quarks of different flavors under weak
  interactions.
\item{} $\langle K^0 | \bar{s}_L \gamma_\mu d \bar{s}_L \gamma_\mu d |
  \overline{K^0} \rangle$ and $\langle B^0 | \bar{b}_L \gamma_\mu d
  \bar{b}_L \gamma_\mu d | \overline{B^0} \rangle$ are the amplitudes
  for the $K^0 \leftrightarrow \overline{K^0}$ and $B^0
  \leftrightarrow \overline{B^0}$ mixing. These play an improtant role
  in the extraction of the CP violation parameters $\epsilon_K, \epsilon_B$ and
  consequently of the Wolfenstein parameters $\rho$ and
  $\eta$~\cite{BurasLH}.
\item{} $\langle \pi \pi | {\mathcal H}_W | \overline{K^0} \rangle$.
  These are the matrix elements of the effective weak Hamiltonian
  describing the weak decays $K \to \pi \pi$. These matrix are needed
  to understand the $\Delta I = 1/2$ rule and in the calculation of
  the CP violating parameter $\epsilon'/\epsilon$~\cite{BurasLH}. .
\end{itemize}

Many simulations testing the efficacy of lattice techniques for
calculating these matrix elements, spectrum of mesons, baryons,
glueballs, the behavior of QCD at finite temperature, and the nature
of the QCD vacuum have been carried out since the pioneering work by
Creutz~\cite{Creutz80} and Wilson~\cite{Wilson79} in the late
seventies.  The current status of these calculations is reported in
the proceedings of the annual conferences on Field Theory on the
Lattice~\cite{LQCDproceedings}.

\section{Computational challenges posed by LQCD}
\label{sec:whyhard}

It is self-evident that given a very precise theory (QCD) and a well
defined method to solve it numerically (Monte Carlo integration of
lattice QCD), the success of the calculations depends on the size of
the statistical and systematic errors compared to the desired
precision.  A brief outline of the systematic errors is as follows.

\begin{itemize}
  
\item {\bf Finite lattice spacing errors}: To make the Feynman path
  integral finite, a discrete space-time grid with spacing $a$ is
  introduced. The continuum theory is recovered in the limit $a \to
  0$.  This limit is taken by extrapolation. At finite $a$ there are
  discretization errors; for example $(f(x+a) - f(x) )/a = \partial_x
  f(x) + a/2 \partial_x^2 f(x) + \ldots$ implies that using such a
  finite difference to approximate a derivative will lead to errors
  proportional to $a$. These errors, as explained by P.
  Lepage~\cite{Lepage}, are being understood and reduced by using (a)
  a generalization of higher order difference schemes that takes
  quantum effects into account and (b) doing the calculations at a
  number of values of $a$ in the range $2 < 1/a < 5$ GeV and then
  extrapolating to $a=0$.
  
\item {\bf Finite Volume errors}: A finite periodic box distorts the
  spectrum of particles. However, it has been determined empirically
  that for $M_{pion} L > 5$ these corrections are exponentially
  suppressed and less than $1\%$. The lattice sizes needed to ensure
  this condition are, for the quenched theory, accessible on computers
  with 100 Gigaflops sustained performance (see the article by A.
  Ukawa\cite{Ukawa}).
  
\item {\bf Quenched Approximation}: The quenched approximation
  distorts the vacuum state of QCD by changing the statistical weight
  of the background gauge configurations. Intuition and model
  calculations suggest that this approximation is valid at the
  $10-20\%$ level for many quantities. In principle, there exist
  algorithms to simulate QCD without making this approximation as
  discussed by A. Kennedy, Ph. de Forcrand and Th.
  Lippert~\cite{Kennedy,Forcrand,Lippert}. However, doing so increases
  the computational time by a factor of $10^3 - 10^5$ depending on the
  value of the quark mass.  I believe that calculations done in the
  next five years using current improved algorithms will be able to
  completely remove this approximation provided we can have access to
  computers with $10-100$ Teraflop sustained performance.
  
\item {\bf Chiral Extrapolation}: The physical up and down quarks are
  very light ($2-8$ MeV).  Iterative algorithms used to solve for
  quark propagators (obtained by inverting the Dirac matrix) exhibit
  critical slowing down that goes as $1/m_{quark}$~\cite{Lippert}.  As
  $m_{quark} \to 0$, the cost for quenched calculations goes up as
  $L^6$ and for unquenched as $L^{8-9}$, where $L$, for fixed $a$, has
  to be increased such that $M_{pion} L \propto \sqrt{m_{quark}} L $
  stays constant. There is good theoretical basis to expect that it
  will be sufficient to calculate physical quantities involving these
  light quarks over the range $m_s - m_s/8$ and then extrapolate to
  $m_u,\ m_d \sim m_s/25$ using relations derived from chiral
  perturbation theory.  Since current quenched calculations go down to
  $m_s/4$ and unquenched to only $m_s$, current data for some quantities 
  has significant uncertainty due to the chiral extrapolation.
  
\item {\bf Discretization of Heavy Quarks}: Heavy quarks, in
  particular the bottom quark with mass $m_b \sim 5$ GeV, has a
  different problem.  It's Compton wavelength is too small compared to
  $a$ for $2 < 1/a < 5$ GeV.  Thus the associated discretization
  errors are large.  Fortunately, novel approaches like the Heavy
  Quark Effective Theory~\cite{HQET} and Non-relativistic
  QCD~\cite{NRQCD91LepageThacker} have shown great promise for
  studying the physics of $b$ quarks.
  
\item {\bf Relation between continuum and lattice quantities}: One of
  the subtleties of QFT is figuring out the correct normalization of
  operators used to calculate matrix elements. These normalization
  constants are essential to relate lattice results to those measured
  in experiments.  In the last five years, the technology to 
  calculate these constants as part of the non-perturbative lattice
  calculations has also been developed and tested~\cite{nonpertZ}.  The
  associated uncertainty has, therefore, been significantly reduced.

\end{itemize}

As of 1999, the two major remaining systematic uncertainties are the
quenched approximation and the discretization of bottom quarks.  A
number of very precise quenched calculations have been done as
discussed
% in articles by N. Christ~\cite{Christ},
%R. Kenway~\cite{Kenway}, 
in the articles of F. Rapuano~\cite{Rapuano} and S. Aoki et
al.~\cite{Aoki} in this volume.  The new generation of
calculations including the effects of dynamical quarks will steadily
tune the sea quark masses towards the physical values of $m_u$, $m_d$,
and $m_s$.  By comparing these results against the precise quenched
estimates, we shall be able to systematically quantify the effects of
vacuum polarization and finally get phenomenologically useful numbers
with all sources of errors estimated.

\section{Why Parallel Computers}
\label{sec:parallel}

From a computational point, numerical algorithms used to simulate
lattice QCD have certain simplifying features.  The calculations are
homogeneous and the interactions are local.  The first feature implies
that exactly the same operations need to be done at
each space-time point.  It is therefore trivial to maintain
synchronization between points.  The second feature implies that only
a few neighboring points are connected at best.  Thus data dependence
is only between small blocks, say $2^4 - 4^4$, or none at all in the
most time consuming part of the calculation (inversion of the Dirac
operator to get the quark propagator).  One can therefore process a
point in each of these blocks simultaneously. In short, the problem is
tailor made for distributed/parallel computing (see articles by
S. G\"usken and Th. Lippert ~\cite{Guesken,Lippert}). 

The 4-dimensional space-time grid is divided into smaller hypercubes,
and the data and calculations needed to process all the sites within
these smaller hypercubes is allocated to each processor.  For example
a $64^4$ lattice can be distributed as $16^4$ sub-lattices on 256
processors set up as a $4^4$ computational grid to
maintain the spatial proximity of the data.  Points internal to this
$16^4$ volume need no information from neighboring processors and get
the maximum throughput allowed by the single processor's speed.
Points on the boundary need communications.  These communications are
again homogeneous. For example each processor needs information from
the processor on the right and simultaneously needs to send
information to the processor on the left.  Furthermore, one almost
always uses periodic boundary conditions which makes this pattern of
communications cyclic on the 4-dimensional grid of processors.
Lastly, one can overlap computations and communications.  The
communications of boundary points can be initiated while the points
internal to the $16^4$ volume are being processed. The only constraint
is the amount of per processor memory to store this
extra data.  This, on todays computers, is not a major hurdle.

The Achilles heel of large distributed computers is inter-node
communication speed and memory bandwidth if commodity
processors are to be used.  In this regard too LQCD calculations
are ideally suited for implementation on parallel computers 
(see the articles of S. Aoki et al., R. Mawhinney, and R.  Tripiccione
in this volume~\cite{Aoki,Mawhinney,Tripiccione}.) One can
examine the worst case senerio of the penalty imposed by
communications in case overlap of communication and computation is not
allowed by either the hardware or the software.  The basic arithmetic
unit in LQCD calculations is the multiplication of a complex $3\times
3$ matrix with a $3\times 1$ complex vector, where the latter may need
communication to a neighboring node.  Thus, one performs $66$ flops
for every $24$ bytes communicated. In practice the situation is even
better as data for only the points on the boundary of the hypercube in
each node need communication. Lastly, global sums and/or global
broadcasts, which are potentially slow, are not done often enough to
be a significant overhead.

Accepting the fact that simulations of LQCD are tailor made for
parallel computing, it should come as no surprise that theoretical
physicists interested in solving QCD have been in the thick of the
development of parallel computers.  The ``cosmic cube'' at Caltech
(1982), the Columbia University special purpose machines (since 1982),
the APE computers of INFN at Rome and Pisa (since 1986), the GF11
supercomputer at IBM (1986-1997), and the Japanese projects at Tsukuba
University (QCDPAX and CP-PACS) are examples of such projects that
have successfully produced Lattice QCD results.  In addition, the
lattice QCD community has also been amongst the first and most
efficient users of commercial supercomputers of all kinds:
vector(CRAYS, ETA10) and/or parallel (Connection Machines, Intel
hypercube and Paragon, IBM SP2, Silicon Graphics Origin 2000).
Consequently, they have contributed to the overall development of high
performance computing by educating a generation of skilled
computational scientists, interacting with computer vendors to help
optimize and test new hardware and software, and providing an
environment and an example that motivated other communities to use
emergent architectures.

%% XXXI SKIP THIS These past successes are discussed in more details by
%% N. Christ~\cite{Christ}, F.
%% Rapuano~\cite{Rapuano}, and A.  Ukawa~\cite{Ukawa} in this volume.
%% Future plans and prospects are summarized by Y.
%% Iwasaki~\cite{Iwasaki}, R.  Mawhinney~\cite{Mawhinney}, and R.
%% Tripiccione~\cite{Tripiccione}, XXXF

\section{Outlook}
\label{sec:Outlook}

Progress in understanding QCD (mechanisms for confinement and chiral
symmetry breaking), verifying QCD (reproducing the hadron spectrum),
calculating strong corrections to weak matrix elements, and
elucidating the behavior of QCD at finite temperature has required a
combination of four things. (i) Better theoretical understanding and
formulation of quantities to calculate: (ii) Better ways of
discretizing QCD as discussed in~\cite{Lepage,Edwards}: (iii) More
efficient numerical algorithms for the generation of background
configurations and quark propagators~\cite{Lippert}: And (iv) higher
performance specifically designed computers and their
exploitation~\cite{Rapuano,Aoki,Mawhinney,Tripiccione}. The field has
witnessed significant developments in all four areas during the last
twenty years. In computing power we have evolved from VAXes ($\sim 1$
Megaflop with 4 Megabytes of RAM) to CP-PACS (600 Gigaflops and 128
Gigabytes of RAM).  As a result we have been able to simulate bigger
lattices and do more precise calculations.  These in turn have
motivated the development of new algorithms and techniques as
discussed in succeeding articles.

I would like to characterize the first twenty years (1979-1998) of
lattice QCD as the taming of the quenched approximation. We have now
calculated many quantities with sufficient precision that the largest
remaining uncertainty is the quenched approximation. Future
calculations will steadily chip away at this last uncontrolled source
of systematic error.  In five years time (year 2004), I anticipate we
shall have access to machines of 10 Teraflops sustained performance,
and will have developed and tested various theoretical and numerical
enhancements that will allow us to obtain the same quality of results
for the full theory. So, it is an exciting coincidence that as we
enter the new millenia, we have a new powerful tool, the computer,
that will allow us to reliably calculate the properties of QCD, a
fundamental theory of nature, from first principles.

\vspace{.2in}
\noindent
\underline{Acknowledgements}

It is a pleasure to thank Nicola Cabibbo, Yoichi Iwasaki, and Klaus
Schilling for inviting me to contribute to this special issue.

%%%%%%%%%%%%%%%%%%%%%%%%%%%%%%%%%%%%%%%%%%%%%%%%%%%%%%%%
% journal and conference list
%%%%%%%%%%%%%%%%%%%%%%%%%%%%%%%%%%%%%%%%%%%%%%%%%%%%%%%%

% Make sure we have href
\ifx\href\undefined\def\href#1#2{{#2}}\fi
% Use macros to link to spires
\def\spireshome{http://www.slac.stanford.edu/cgi-bin/spiface/find/hep/www?FORMAT=WWW&}
{\catcode`\%=12
  \xdef\spiresjournal#1#2#3{\noexpand\protect\noexpand\href{\spireshome
                          rawcmd=find+journal+#1%2C+#2%2C+#3}}
\xdef\spireseprint#1#2{\noexpand\protect\noexpand\href{\spireshome rawcmd=find+eprint+#1%2F#2}}
    \xdef\spiresreport#1{\noexpand\protect\noexpand\href{\spireshome
        rawcmd=find+rept+#1}}
    \xdef\spireskey#1{\noexpand\protect\noexpand\href{\spireshome
        key=#1}} } \def\eprint#1#2{\spireseprint{#1}{#2}{#1/#2}}
  \def\report#1{\spiresreport{#1}{#1}} \def\nohref{}

% Something to help define the journals
\def\putpaper{\edef\refpage{\the\count0}%
              \def\nohref{}%
              {\def\ 
                {+}\def\nohref##1{}\edef\temp{\noexpand\spiresjournal
                  {\journalname}{\volume}{\refpage}}\expandafter}\temp
              {\sfcode`\.=1000{\journalname} {\bf \volume} (\refyear)
                \refpage}\egroup}
\def\putpage{\edef\refpage{\the\count0}%
              \def\nohref{}%
              {\def\ 
                {+}\def\nohref##1{}\edef\temp{\noexpand\spiresjournal
                  {\journalname}{\volume}{\refpage}}\expandafter}\temp
              {\refpage}\egroup} \def\dojournal#1#2
            (#3){\def\journalname{#1}\def\volume{#2}\def\refyear
              {#3}\afterassignment\putpaper\bgroup\count0=}
            \def\morepage{\afterassignment\putpage\bgroup\count0=}
            \def\supresslink{\def\spiresjournal##1##2##3{}}

% Okay, now the list of journals
            \def\APNY#1{\dojournal{Ann.\ Phys.\ \nohref{(N.\ 
                  Y.)}}{#1}} \def\CMP#1{\dojournal{Comm.\ Math.\ 
                Phys.}{#1}} \def\IJMPC#1{\dojournal{Int.\ J.\ Mod.\ 
                Phys.}{C#1}} \def\IJMPE#1{\dojournal{Int.\ J.\ Mod.\ 
                Phys.}{E#1}} \def\JAP#1{\dojournal{J.\ App.\ 
                Phys.}{#1}}
            
            \def\MPA#1{\dojournal{Mod.\ Phys.\ Lett.}{A#1}}
            \def\MPLA#1{\dojournal{Mod.\ Phys.\ Lett.}{A#1}}
            \def\NP#1{\dojournal{Nucl.\ Phys.}{B#1}}
            \def\NPA#1{\dojournal{Nucl.\ Phys.}{A#1}}
            \def\NPB#1{\dojournal{Nucl.\ Phys.}{B#1}}
            \def\NPBPS#1{\dojournal{Nucl.\ Phys.\ \nohref(Proc.\ 
                Suppl.\nohref)}{\nohref B#1}}
            \def\NPAPS#1{\dojournal{Nucl.\ Phys.\ \nohref(Proc.\ 
                Suppl.\nohref)}{\nohref A#1}}
            \def\NC#1{\dojournal{Nuovo Cimento }{#1}}
            \def\PRL#1{\dojournal{Phys.\ Rev.\ Lett.}{#1}}
            \def\PR#1{\dojournal{Phys.\ Rev.}{#1}}
            \def\PRep#1{\dojournal{Phys.\ Rep.}{#1}}
            \def\PRB#1{\dojournal{Phys.\ Rev.}{B#1}}
            \def\PRC#1{\dojournal{Phys.\ Rev.}{C#1}}
            \def\PRD#1{\dojournal{Phys.\ Rev.}{D#1}}
            \def\PRE#1{\dojournal{Phys.\ Rev.}{E#1}}
            \def\PL#1{\dojournal{Phys.\ Lett.}{#1B}}
            \def\PLA#1{\dojournal{Phys.\ Lett.}{#1A}}
            \def\PLB#1{\dojournal{Phys.\ Lett.}{#1B}}
            \def\RMP#1{\dojournal{Rev.\ Mod.\ Phys.}{#1}}
            \def\PREP#1{\dojournal{Phys.\ Rep.}{#1}}
            \def\ZEITC#1{\dojournal{Z.\ Phys.}{C#1}}
            \def\ZPC#1{\dojournal{Z.\ Phys.}{C#1}}
            
            \def\ie{{\sl i.e.}} \def\etal{{\it et al.}} \def\etc{{\it
                etc.}} \def\ibid{{\it ibid}}

%%%%%%%%%%%%%%%%%%%%%%%%%%%%%%%%%%%%%%%%%%%%%%%%%%%%%%%%%%%%%%%%%%
%%%%%%%%%%%%%%%%%%%%%%%%%%%%%%%%%%%%%%%%%%%%%%%%%%%%%%%%%%%%%%%%%%

\end{document}